# Positronium impact ionization of Alkali atoms

D. Ghosh[1] and C. Sinha[2]

[1]Michael Madhusudan memorial College, Durgapur, Burdwan, West Bengal, India

[2]Theoretical Physics Dept, Indian Association for the cultivation of science, Kolkata, West Bengal, India

e- mail: dipalighoshwb@gmail.com

**Abstract:** Target ionization processes of alkali atoms by Positronium impact are investigated. Calculations are performed in the frame work of model potential formalism using the Coulomb distorted eikonal approximation. Interesting qualitative features are noted both in the scattered Ps and the ejected electron distributions in triple differential as well as double differential levels of the collision cross sections.

1. Introduction:

Alkali atoms as targets in atomic collisions have been considered as a large numbers of experimental and theoretical studies over the years, as they can be addressed both theoretically and experimentally relatively easily such that detailed comparison is possible [1]. Again collisions with alkali- metal atoms are of interest for application in the design of laser systems in VUV regime [2] and for diagnostics of (fusion) plasmas and their impurities [3, 4]. Further, from the theoretical point of view, the shell structure of alkalis are interesting in the sense that the quasi-one electron models of the loosely bound outermost electron and a stationary effective potential due to frozen inner shell electrons are the appealing features of the alkali atoms involved in collision processes. On the other hand, use of positronium as the projectile becomes interesting as new experimental techniques and theoretical methods are enabling increasingly stringent tests of the understanding of basic atomic and molecular collision phenomena as well as of fundamental antiparticle-matter interactions. In this paper a comparative study of Ps impact ionization of two alkali atoms are performed. Such collision processes find interest as the targets are alkali atoms having simple structures, low ionization potentials and large polarizabilities where as the projectile being the simplest particle- antiparticle system.

From the theoretical perspective, single ionization process by Ps impact, even of the simplest hydrogenic target is a bit difficult task [5] as it becomes a four body problem. The complexity mainly arises due to the internal degrees of freedom of the projectile Ps which must be taken into account. However the direct Coulomb interaction between the Ps and the atomic target is very much smaller as compared to that arising from the electron exchange effect between them [6]. Again the calculation of this exchange process is rather difficult since it involves electron swapping between two different centres, the target and the Ps though the electron exchange effect seems to be not the main driving force for the target ionization process. Therefore, our main task is to develop suitable mathematical tools necessary for solving the many body Schrodinger equation that appears in this particular single ionization process. Various approximation models like close coupling [7 -9], R- matrix theory [10-11], different variational methods [12- 14] were developed for solving the Ps impact elastic and inelastic scattering processes. From such investigations one can make a comparative study of the different theoretical models with relative merits and demerits and their agreements with the available experimental data. The suitability of different mathematical models depends on the collision partners, particular collisional channel and on the energy regime concerned. Close coupling (CC) methods are proved to e be quite successful in the lower incident energy regime while the CDW models are supposed to be more suitable with reasonable accuracy at comparatively moderate and higher incident energies for which the CC methods become increasingly difficult with energies to compute with.

The recent experimental and theoretical results of Brawley et al [ 15 ] show that the total cross sections of atoms with Ps as projectile is unexpectedly close to that of a bare electron projectile moving at the same velocity. This findings motivated us to study theoretically the target ionization process of alkali atoms by Ps impact and to compare the results with the corresponding electron impact ionization results. The present model is based on the frame work of Coulomb Modified Distorted Wave Approximation (CMDA). We have calculated both the triple differential cross sections (TDCS) and the double differential cross sections (DDCS) and have tried to compare the findings with the existing theoretical and experimental results [16- 18] of electron impact ionization of Na atom.

The basic difference between the electron impact and the Ps impact ionization lies in the fact that in the latter case, both the projectile and the target are composite objects having an internal structure and as such the dynamics demands evaluation of multicenter integrals occurring in the transition matrix elements which are quite difficult and time consuming. The present study of target ionization ( by Ps impact) is different from that of the single ionization of the target atom/ion by positron or electron impact and as such the present TDCS additionally carries the information about the influence of the Ps on the ejected electron distributions as the TDCS of the ejected electron varies with both the energy and the angle of the scattered Ps. The inclusion of the exchange effect between the projectile electron and the target electron in the final channel would lead to formidable difficulties in the present prescription.

The present problem addresses the theoretical study of the dynamics of target inelastic process, e.g., single ionization of the target ( Na and K atom) , both being initially in their ground states.

$$e^+e\ (1s) + \ X(1s) \rightarrow e^+e(1s) + X^+ + e \tag{1}$$

where $X = Na, K$

Since the initial components of this interaction are both composite bodies, the theoretical prescription of such a process is rather complicated and as such one has to resort to some simplifying assumptions for the theoretical models of such a many-body reaction process. The present calculation is performed in the framework of post collisional Coulomb distorted eikonal approximation taking account of the proper asymptotic boundary condition of the ejected electron in final channel, which is one of the most important criteria for a reliable estimate of the ionization cross-sections.

**2. Theory:**

The prior form of the ionization amplitude for the aforesaid process (1) is given as:

$$T_{if}^{prior} = -\frac{\mu_f}{2\pi} \left\langle \Psi_f^-(\vec{r}_1, \vec{r}_2, \vec{r}_3) \left| V_i \right| \psi_i(\vec{r}_1, \vec{r}_2, \vec{r}_3) \right\rangle \tag{2}$$

The initial asymptotic wave function $\psi_i$ in equation (2) is chosen as

$$\psi_i = \phi_{Ps}(|\vec{r}_1 - \vec{r}_2|) \, e^{i\vec{k}_i \cdot \vec{R}} \, \phi_T(\vec{r}_3) \tag{3a}$$

where $\vec{R} = (\vec{r}_1 + \vec{r}_2)/2$ and $k_i$ is the initial momentum of the Ps atom with respect to the target nucleus. The ground state wave function of the Ps atom

$$\phi_{Ps}(|\vec{r}_1 - \vec{r}_2|) = N_{1s} \exp(-\lambda_i r_{12}) \tag{3b}$$

with $N_{1s} = \lambda_i^{3/2}/\sqrt{\pi}$ and $\lambda_i = 1/2$. The ground state wave function of the target sodium or potassium atom is chosen in the form of a simple hydrogenic orbital as

$$\phi_T(r_3) = N_T \exp(-\lambda_T r_3) \tag{4}$$

The value of $\lambda_T$ is taken from the work of Hart and Goodfriend [19] and $N_T = \lambda_T^{3/2}/\sqrt{\pi}$

The complexity of working with many electron atom have been circumvented in different theoretical investigation [20- 25] by considering the model potential [26, 27], where the effect of the core electrons have not been considered explicitly. The model potential of the alkali atoms initiates the multi- electron core interaction with the single valence electron by an analytic modification of the Coulomb potential. In the present calculation $V_i$ is the initial channel perturbation not diagonalized in the initial state is chosen as model potential following the work of Schweizer et al [28] given by,

$$V_i = \frac{1}{r_1} - \frac{1}{r_2} - \frac{1}{r_{13}} + \frac{1}{r_{23}} + \frac{N}{r_1} \exp(-a_1 r_1) - \frac{N}{r_2} \exp(-a_1 r_2) + a_2 \exp(-a_3 r_1) - a_2 \exp(-a_3 r_2) \tag{5}$$

$\vec{r}_1$, $\vec{r}_2$ and $\vec{r}_3$ in eqn.(2) are the position vectors of the positron and the electron of the Ps and the bound electron of the target atom (Na and K) respectively, with respect to the target nucleus; N= 10 and 18 for Na and K respectively and $\mu_f$=2. The values of $a_1$, $a_2$ and $a_3$ are taken from the work of Schweizer et al [28].

Here, $\vec{r}_{13} = \vec{r}_1 - \vec{r}_3$ and $\vec{r}_{23} = \vec{r}_2 - \vec{r}_3$.

The wave-function $\Psi_f^-$ satisfies the incoming wave boundary condition. The corresponding Schrodinger equation is given by,

$$(H - E)\Psi^\pm = 0 \qquad (6)$$

where the full Hamiltonian of the system is given by,

$$H = -\frac{\nabla_R^2}{2\mu_i} - \frac{\nabla_{12}^2}{2\mu_{ps}} - \frac{\nabla_3^2}{2} - \frac{1}{r_{12}} + \frac{Z_t}{r_1} - \frac{Z_t}{r_2} - \frac{Z_t}{r_3} - \frac{1}{r_{13}} + \frac{1}{r_{23}}$$

$$+ \frac{N}{r_1}\exp(-a_1 r_1) - \frac{N}{r_2}\exp(-a_1 r_2) + a_2 \exp(-a_3 r_1) - a_2 \exp(-a_3 r_2)$$

where $\mu_i$ and $\mu_{Ps}$ are 2 and 1/2 respectively.

In the present work we have adopted the prior version of the transition matrix (eqn.(2)) which is supposed to be more suitable for an ionization process [29- 32]. Equation (6) concerning a four body problem could not be solved exactly and as such one has to resort to some simplifying assumptions. The final state wave function $\Psi^-_f$ ( eqn.(2)) involving two bound particles (Ps) and one continuum particle is approximated by the following ansatz in the framework of Coulomb – eikonal approximation [ 31- 34] :

$$\Psi_f^-(\vec{r_1}, \vec{r_2}, \vec{r_3}) = N_{1s} \exp(-\lambda_f r_{12}) N_3 (2\pi)^{-3/2} e^{i\vec{k_3}.\vec{r_3}} {}_1F_1(-i\alpha_3, 1, -i(k_3 r_3 + \vec{k_3}.\vec{r_3}))$$

$$e^{i\vec{k_f}.\vec{R}} \exp\left\{i\eta_f \int_z^\infty \left(\frac{1}{r_1} - \frac{1}{r_2}\right)dz'\right\} \qquad (7)$$

where

$N_3 = \exp\left(\frac{\pi\alpha_3}{2}\right)\Gamma(1-i\alpha_3)$ with $\alpha_3 = -\frac{1}{k_3}$, $\eta_f = \frac{1}{k_f}$ ; and $\lambda_f = \lambda_i = 1/2$; since the Ps remains in the ground state in final channel.

$\vec{k_3}$ and $\vec{k_f}$ are the final momentum of the ejected electron and the positronium respectively. Equation (7) satisfies the incoming wave boundary condition which is one of the essential criteria for a reliable estimate of an ionization process.

The two centre effect on the electron of the Ps due to its parent ion ($e^+$) and the screened target ion is implicit in eqn. (7). Since in the final channel the ejected electron from the target is in the long range Coulomb field of the residual target ion ( $Na^+$, $K^+$), this interaction is incorporated in eqn. (7) . The justification of the present ansatz for the approximate wave function $\Psi_f^-$ can be given as follows. The confluent hypergeometric function ($_1F_1$) arises because of the continuum wave function of the ejected electron in the field of its parent target ion. The strong interactions between the target nucleus and the two components of the incident particle ( e & $e^+$ of Ps ) are taken into account by the two eikonal factors in the final channel. In order to avoid the complexity in the analytical calculations, we have neglected the higher order interactions between the $e^+$ / e of the Ps and the target electron and have mainly concentrated on the ionization of the target; this interaction being considered through the perturbation interaction in the initial channel.

In view of equations ( 2 – 7 ) , we obtain the target ionization amplitude (direct ) for the process (1) as

$$T_{if}^{direct} \equiv -\frac{\mu_f}{2\pi} \iiint N_3^* N_T (2\pi)^{-3/2} \exp(-\lambda_T \vec{r}_3) e^{i\vec{k}_i \cdot \vec{R}} (N_{1s})^2 \exp(-\lambda r_{12})$$

$$(\frac{z_t}{r_1} - \frac{z_t}{r_2} - \frac{1}{r_{13}} + \frac{1}{r_{23}} + \frac{N}{r_1} \exp(-a_1 r_1) - \frac{N}{r_2} \exp(-a_1 r_2) + a_2 \exp(-a_3 r_1) - a_2 \exp(-a_3 r_2))$$

$$e^{-i\vec{k}_3 \cdot \vec{r}_3} e^{-i\vec{k}_f \cdot \vec{R}} (r_1 + z_1)^{i\eta_f} (r_2 + z_2)^{-i\eta_f} {}_1F_1(i\alpha_3, 1, i(k_3 r_3 + \vec{k}_3 \cdot \vec{r}_3)) \, d\vec{r}_1 \, d\vec{r}_2 \, d\vec{r}_3$$

(8)

Where $\lambda = \lambda_i + \lambda_f$. After much analytical reduction [35- 38] the target ionization amplitudes $T_{if}$ in equation (8) is finally reduced to a three dimensional numerical integral. The triple differential cross sections (TDCS) [36] is given by

$$\frac{d^3\sigma}{dE_3 d\Omega_f d\Omega_3} = \frac{k_f k_3}{k_i} |T_{if}|^2 \qquad (9)$$

and the double differential cross sections (DDCS) i.e., $\frac{d^2\sigma}{dE_3 d\Omega_f}$ are obtained by integrating over the solid angle $d\Omega_3$.

It may be mentioned in this context that due to the principle of detailed balance , the transition amplitude obtained from the post and prior forms should in principle, be the same if the exact scattering wave function in the initial or final channel ($\Psi_i^+, \Psi_f^-$) could be used, which for a four body problem is a formidable task. In the case of approximate wave functions, the afore said two forms might not lead to identical results giving rise to some post – prior discrepancy. However, in the case of simple First Born Approximation (FBA) where the initial or final scattering states are represented by the corresponding asymptotic wave functions, there should not be any post – prior discrepancy.

## 3. Results and Discussion

The TDCS and the DDCS results are computed for target ionization of the alkali atoms, Na and K by Ps impact. For the single ionization, the threshold energy is determined by $E_{th} = E^{1s}_{(Na,K\ )}$. Since the present study is made in coplanar geometry, i.e., $\vec{k}_i$, $\vec{k}_f$ and $\vec{k}_3$ all being in the same plane, the azimuthal angles $\phi_f$, $\phi_f$ and $\phi_3$ can assume values $0^0$ and $180^0$. For the TDCS curves, we have adopted the following conventions for the ejected angles $(\theta_3, \phi_3)$: $(\theta_3, 0^0)$ we have denoted by $-|\theta_3|$ (recoil region) while the angles $(\theta_3, 180°)$ are plotted as $|\theta_3|$ binary region.

Figure 1 exhibits the angular distributions (TDCS in atomic unit ( a.u.) ) of the ejected electron ($\theta_3$) for the sodium target. To make a comparison with the results of electron - impact single ionization of sodium by Armstrong et al [18], the incident energy ($E_i$) is kept fixed at 11.138 eV for different scattering angles of the incident Ps ($\theta_f = 0^0$, $30^0$, $60^0$, $90^0$, $120^0$ and $150^0$). We have compared the TDCS for both equal velocity and equal energy sharing between the ejected electron and the scattered positronium. It should also be pointed out here that the mass of the Ps being double to that of the electron, we have considered the Ps ejected energy to be twice the ejected energy of the electron to keep pace between the ejected electron and the Ps in the velocity space.

As is evident from Figs. 1(a) to 1(f), the TDCS shows single peak with some associated hump like structures. For lower scattering angle, forward ejection is preferred having higher values of cross sections and for this particular incident energy, the velocity matching between the electron and the Ps dominates the equal energy sharing between them. Comparison between electron and Ps impact ionization shows that for smaller values of the scattering angle, the Ps impact ionization takes over the electron impact ionization where as for higher scattering angle, the electron impact ionization dominates over the Ps one.

Figure 2 illustrates the Ps impact fully differential equal energy sharing ionization cross sections in the symmetric geometry. The figure reveals that for higher incident energy ( 35.138 eV; fig. 2a), a binary peak is prominent but at lower incident energy (20.138 eV; fig. 2b), no such prominent peak is noted but comparatively much higher magnitude (~ 5 times ) arises at extreme forward ejections. Again, it shows that at lower incident energy, velocity matching between the electron and Ps is more preferred where as for higher incident energy equal energy sharing between them predominates.

Figures 3(a) to 3(d) show similar ejected electron distributions (TDCS ; as in Fig. 1) for potassium atom corresponding to four different kinematics keeping $\theta_f = 0^0$. From these figures it is clear that a single peak associated with some shoulder like structures arises for higher incident energy and equal velocity whereas for equal energy sharing a backward hump grows.

Figures 4(a) to 4(d) represent the TDCS in symmetric geometry (as in Fig. 2) for potassium atom. For different incident energies, kinematics are chosen for both equal energy sharing and velocity matching of the ejected electron and the scattered Ps. Like sodium atom, for potassium atom also the symmetric geometry TDCS shows preference for equal energy sharing at higher incident energy. Qualitative comparison between Ps and electron impact ionization reveals that a single peak arises in the binary side at higher incident energy for both the cases while at lower incident energy, the peak is not formed fully for Ps impact ionization.

The next figures 5(a) and 5(b) exhibit the double differential cross sections (DDCS) with respect to the scattered Ps angle ($\theta_f$) for the Ps - Na/ K system. Figs. 5(a) and 5(b) depict the Ps distributions for the fixed incident energy at 25 eV keeping the ejected electron energy fixed at 5 eV and 10 eV respectively. It is revealed that the over all distributions of the DDCS for both Sodium and Potassium are same though the magnitude differs depending on the kinematics.

Conclusion:

1. For some particular kinematics, qualitative similarities between the electron and the positronium impact ionization of the alkali atoms ( Na,K ) are revealed, in conformity with experiment [15] while for some other kinematics, both qualitative and quantitative discrepancies are prominent between the two cases.

2. Study of both the Equal energy sharing and the velocity matching between the ejected electron and the positronium shows that both of these kinematics depend upon the incident energy as well as on the collision geometry.

3. The DDCS with respect to the scattered Ps angle shows similar qualitative behaviour for both the alkali atoms sodium and potassium. Regarding the magnitude of the DDCS, it can be inferred that for lower and higher ejection energy, Na and K atom dominates each other respectively.

**Acknowledgements**

This work is supported by the University Grants Commission, India under the grant No. PSW-026/ 13-14 (ERO) ID No. WB1/053.

**References:**

[1] Zapukhlyak M, Kirchner T, Ludde H J, Knoop S, Morgenstern R and Hoekstra R 2005 *J. Phys. B : At. Mol. Opt. Phys.* **38** 2353

[2] Aumayr F and Winter H 1987 *J. Phys. B : At. Mol. Opt. Phys.* **20** L803

[3] Horvath G, Schweinzer J, Winter H and Aumayr F 1996 *Phys. Rev. A* **54** 3022

[4] Ebel F, Salzborn E 1987 *J. Phys. B : At. Mol. Opt. Phys.* **20** 4531

[5] Ray H 2004 *Pramana* **63** No. 5 1063

[6] Blackwood J E, Campbell C P, McAlinden Mary T and Walters H R J 1999 *Physical Review A* **60** 4454

[7] Ray H and Ghosh A S 1998 *J. Phys. B : At. Mol. Opt. Phys.* **31** 4427

[8] Basu A, Sinha P K and Ghosh A S 2000 *Physical Review A* **63** 012502

[9] Biswas P K and Darewych J W 2002 *Nucl. Istrum.Methods Phys Res. B* **192** 138

[10] Campbell C P, McAlinden M T, MacDonald F G R S and Walters H R J 1998 *Phys Rev Lett*. **80** 5097

[11] Blackwood J E, McAlinden M T and Walters H R J 2002 *Phys. Rev. A* **65** 032517

[12] Ivanov I A, Mitroy J and Varga K 2001 *Phys Rev. Lett*. **87** 063201

[13] Reeth P V and Humberston J W 2003 *J Phys B* **36** 1923

[14] Chiesa S, Mella M and Morosi G 2002 *Phys. Rev. A* **66** 042502

[15] Brawley S J, Armitage S, Beale J, Leslie DE, Williams A I and Laricchia G 2010 *Science* **330** 789

[16] Bray I, Fursa D V and Stelbovics A T 2009 *J Phys B: Conference Series* **185** 012003

[17] Wang Y, Jiao L, Zhou Y 2012 *Physics Letters A* **376** 2122

[18] Armstrong G S J, Colgan J and Pindzola M S 2013 *Phys. Rev. A* **88** 042713

[19] Hart G A and Goodfriend P L 1970 *The Journal of Chemical Physics* **53** 448

[20] Guha S and Mandal P 1980 *J. Phys. B : At. Mol. Opt. Phys.* **13** 1919

[21] Sinha C, Guha S and Sil N C 1982 *J. Phys. B : At. Mol. Opt. Phys.* **15** 1759

[22] Cavaliere P, FerrenteG and Montes B M 1975 *Chem. Phys. Lett*. **36** 583

[23] Bordonaro G, Ferrante G, Zarcone M and Cavaliere P 1976 *Nuovo Cim. B* **35** 349

[24] Ferrante G, Cascio L L and Zarcone M 1978 *Nuovo Cim B* **44** 99

[25] Montes B M, Cavaliere P and Ferrante G 1975 *Chem. Phys. Lett*. **32** 469

[26] Cavaliere Pand Ferrante G 1973 *Nuovo Cim B* **14** 127

[27] Hannsen J, McCarrol R and Valiron P 1979 *J Phys B* **12** 899

[28] Schweizer W, Fabbinder P and Gonzalez-Ferez R 1999 *Atomic Data and Nuclear Data Tables* **72** No. 1 133

[29] Brauner M, Briggs J Klar H 1989 *J. Phys. B* **22** 2265

[30] Biswas R and Sinha C 1994 *Phys. Rev. A* **50** 354

[31] Roy S, Ghosh D and Sinha C 2005 *J. Phys. B* **38** 2145

[32] Roy S and Sinha C 2009 *Phys. Rev. A* **80** 022713

[33] Fried H M, Kang K and Mckeller B H J 1983 *Phys. Rev. A* **28** 738

[34] Ghosh D, Mukhopadhyay S and Sinha C 2013 *Eur. Phys. J. D* **67** 85

[35] Sinha C and Sil N C 1978 *J. Phys. B* **11** L333

[36] Nath B and Sinha C 2000 *J. Phys. B* **33** 5525

[37] Ghosh D and Sinha C 2004 *Phys. Rev. A* **69** 052717

[38] Avaldi L, Camilloni P and Stefani G 1990 *Phys. Rev. A* **41** 134

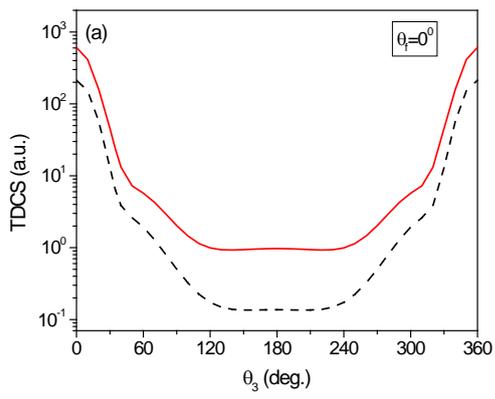
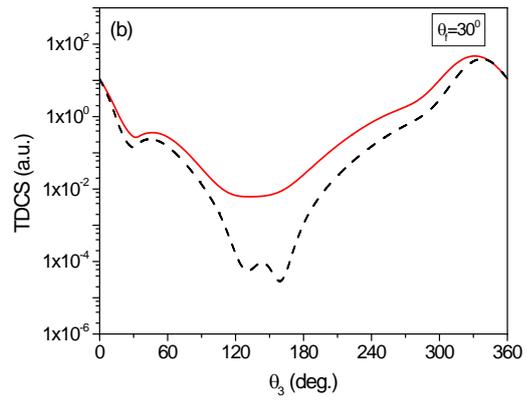
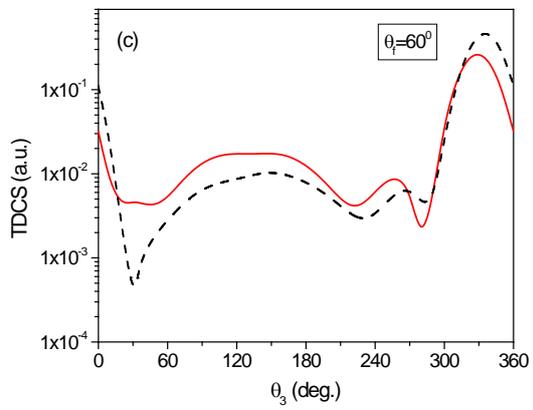
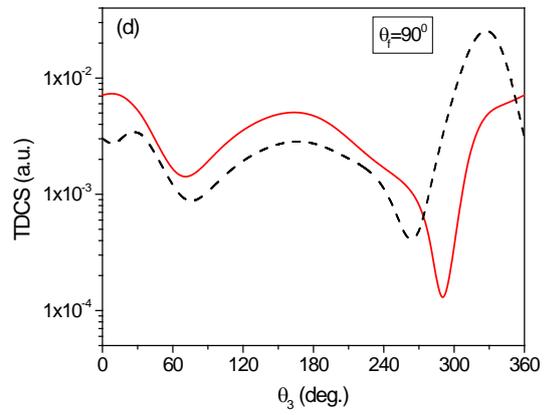

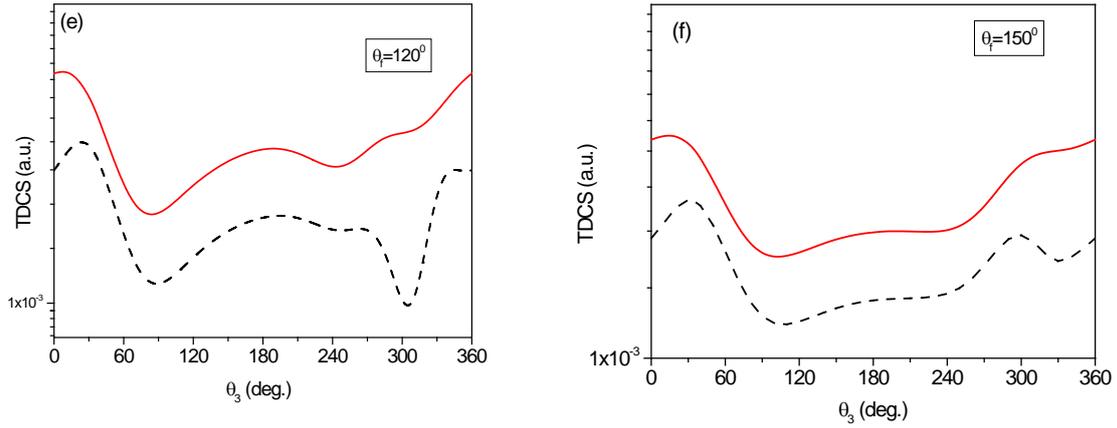

Figure 1. (color online) Triple differential cross section for Ps impact ionization of Sodium for different values of scattering angle at an incident energy 11.138 eV in the coplanar asymmetric geometry ($\phi_3 = \phi_f = 0^0$). The dashed curve represents equal energy sharing ($E_3=E_f= 3$ eV) and the solid curve represents for equal velocity of the ejected electron and scattered positronium ($E_3=2$ eV, $E_f=4$ eV).

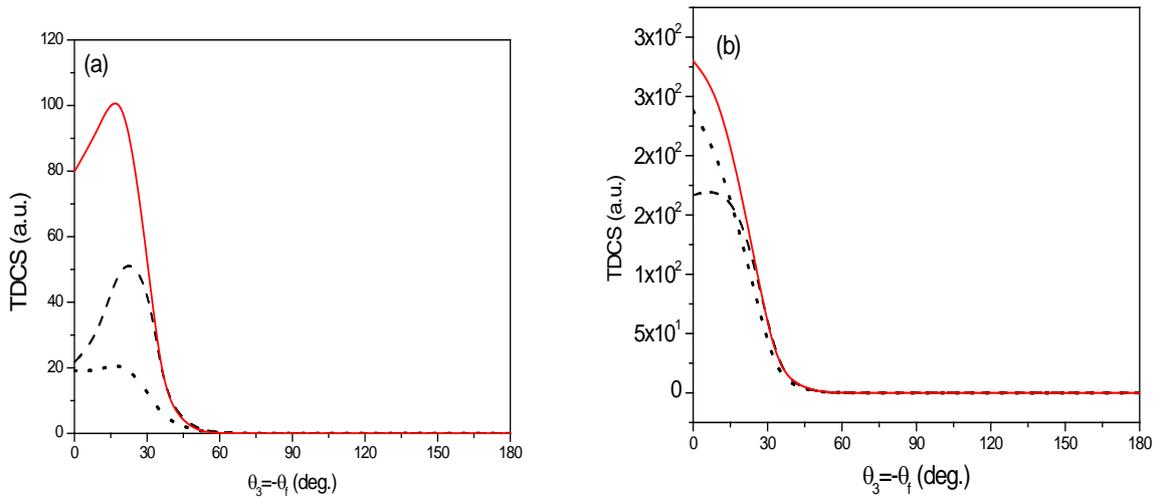

Figure 2. (color online) Triple differential cross section for Ps impact ionization of Sodium in the coplanar symmetric geometry ($\theta_3 = -\theta_f$). In fig. 2a the dashed curve represents the kinematics $E_i= 35.1$ eV, $E_3=E_f= 15$ eV and the dotted curve represents the kinematics $E_i= 35.138$ eV, $E_3= 10$ eV and $E_f= 20$ eV. The solid curve represents equal energy sharing for a lower incident energy $E_i= 25.138$ eV $E_3=E_f= 10$ eV. In fig. 2b the dashed curve represents the kinematics $E_i= 20.138$ eV, $E_3=E_f= 7.5$ eV and the dotted curve represents the kinematics $E_i= 20.138$ eV, $E_3= 5$ eV and $E_f= 10$ eV. The solid curve represents equal energy sharing for a lower incident energy $E_i= 15.1$ eV $E_3=E_f= 5$ eV.

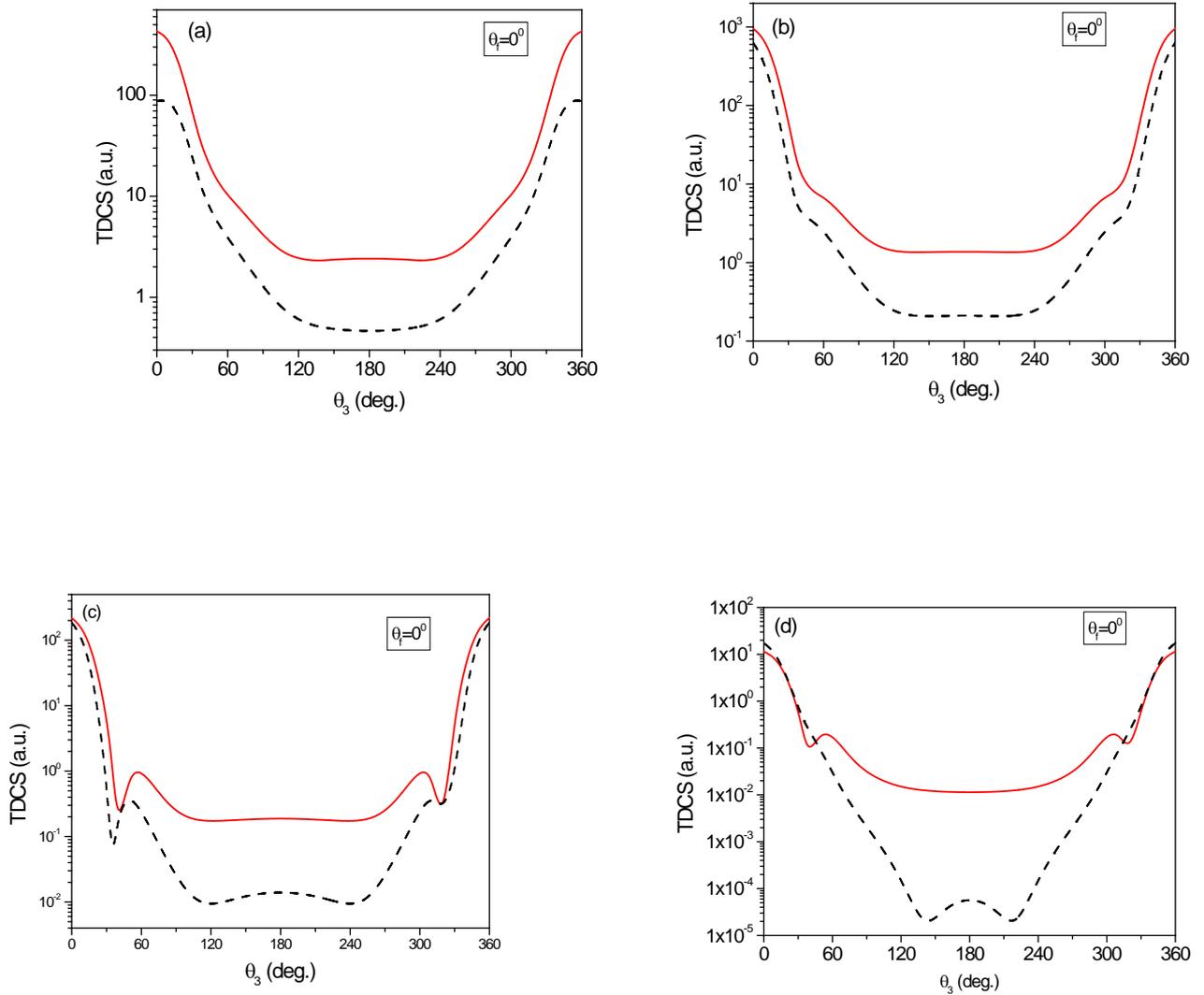

Figure 3. (color online) Triple differential cross section for Ps impact ionization of Potassium for the fixed scattering angle $0^0$ for different incident energies in the coplanar asymmetric geometry . In fig. 3a the incident energy is kept fixed at 7.341 eV. The dashed curve represents equal energy sharing ($E_3=E_f=$ 1.5 eV) and the solid curve represents for equal velocity of the ejected electron and scattered positronium ($E_3=1$ eV, $E_f=2$eV). Fig. 3b is same as fig. 3a but the incident energy is kept fixed at 10.341 eV. For the dashed curve ($E_3=E_f=$ 3 eV) and the for solid curve ($E_3=2$, $E_f=$ 4 eV). Fig. 3c shows the curve having incident energy 19.341 eV. Here the dashed curve represents ($E_3=E_f=$ 7.5 eV)and solid curve represents ($E_3=5$, $E_f=$ 10 eV). Fig. 3d shows the curve having incident energy 34.341 eV. Here the dashed curve represents ($E_3=E_f=$ 15 eV)and solid curve represents ($E_3=10$, $E_f=$ 20 eV).

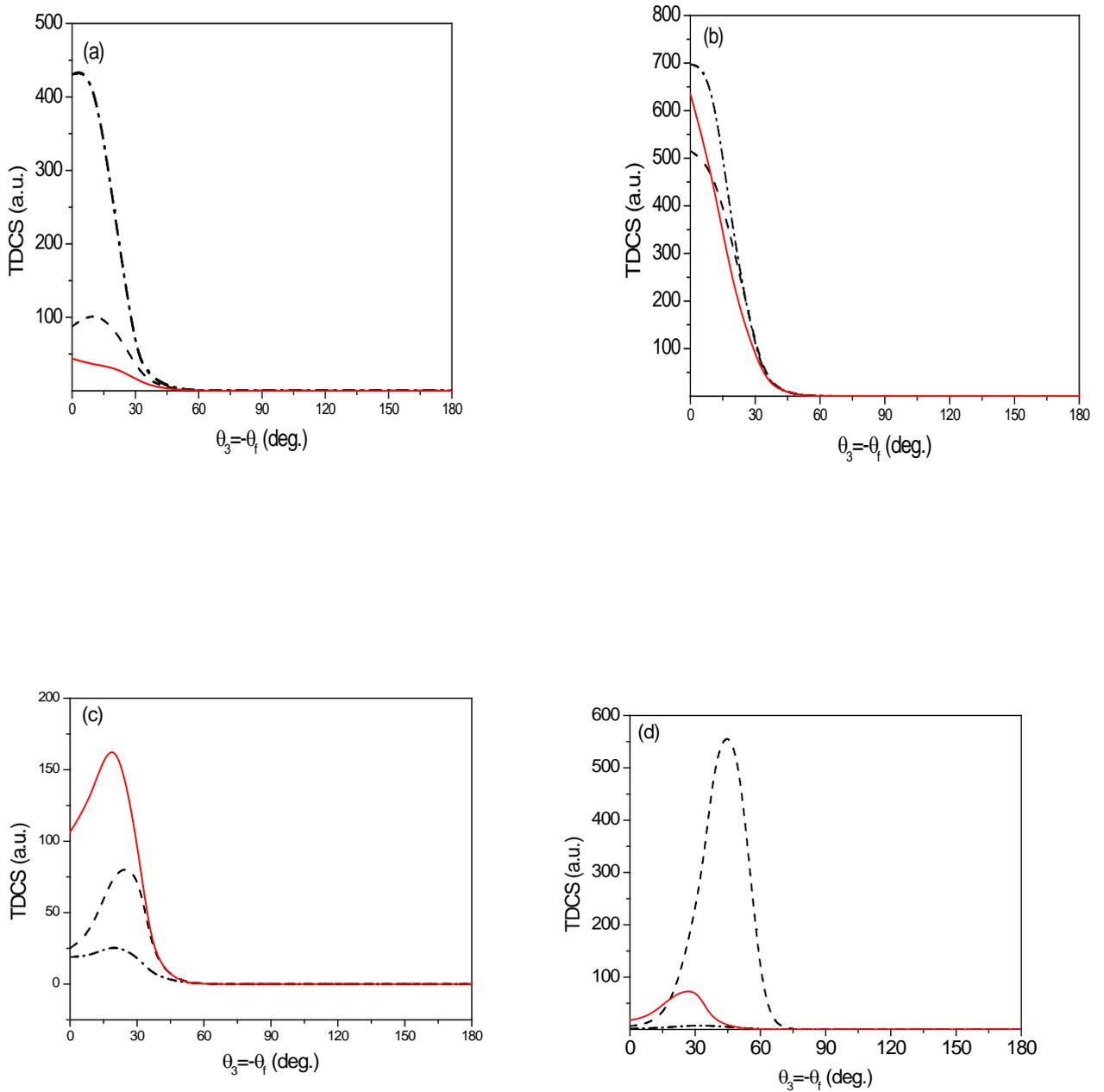

Figure 4. (color online) Triple differential cross section for Ps impact ionization of Potassium in the coplanar symmetric geometry ($\theta_3 = -\theta_f$). In fig. 4a the dashed curve represents the kinematics $E_i$= 7.341 eV, $E_3$=$E_f$= 1.5 eV  and for same incident energy dash-dotted curve represents the kinematics $E_3$= 1 eV and $E_f$= 2 eV.  The solid curve represents equal energy sharing  for a lower incident energy $E_i$= 6.341 eV $E_3$=$E_f$= 1 eV. In fig. 4b the dashed curve represents the kinematics $E_i$= 13.341 eV, $E_3$=$E_f$= 4.5 eV  and the dash- dotted curve represents the kinematics $E_i$= 13.341 eV, $E_3$= 3 eV and $E_f$= 6 eV. The solid curve represents equal energy sharing  for a lower incident energy $E_i$= 10 eV $E_3$=$E_f$= 3 eV. In fig. 4c the dashed curve represents the kinematics $E_i$= 31.341 eV, $E_3$=$E_f$= 13.5 eV   and  for same incident energy dash-dotted curve represents the kinematics $E_3$= 9 eV and $E_f$= 18 eV.  The solid curve represents equal energy sharing  for a lower incident energy $E_i$= 22.341 eV $E_3$=$E_f$= 9 eV. In. In fig. 4d the dashed curve represents the kinematics $E_i$= 49.341 eV, $E_3$=$E_f$= 22.5 eV   and for same incident

energy dash-dotted curve represents the kinematics $E_3$= 15 eV and $E_f$= 30 eV. The solid curve represents equal energy sharing for a lower incident energy $E_i$= 34.341 eV $E_3$=$E_f$= 15 eV.

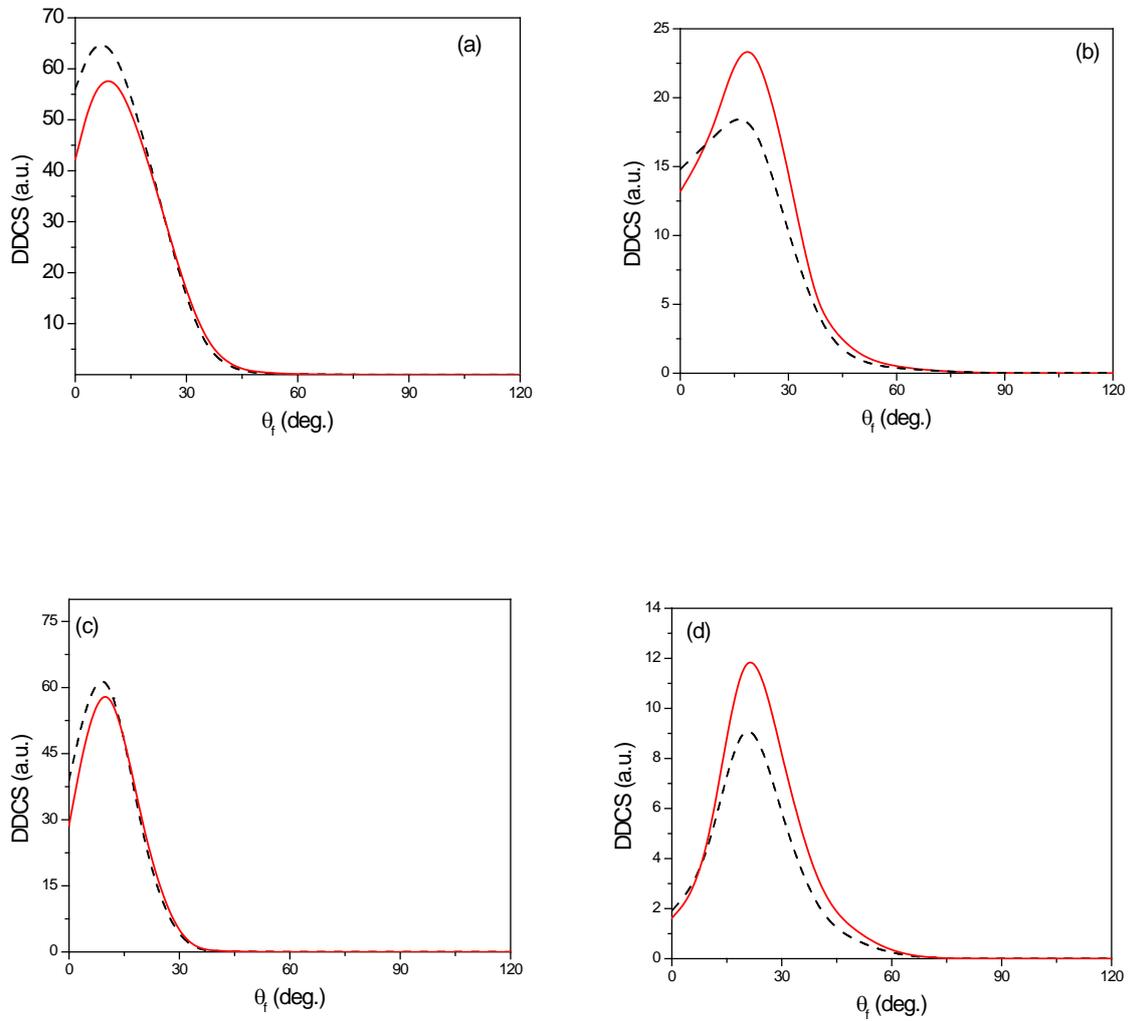

Figure 5. (color online) Double differential cross section for Ps impact ionization of Sodium (dashed curve) and Potassium (solid curve) in different asymmetric kinematics. In fig. 5a, $E_i$= 25 eV, $E_3$=5 eV , in fig. 5b , $E_i$= 25 eV, $E_3$=10 eV in fig. 5c $E_i$= 45 eV, $E_3$=5 eV and in Fig. 5d $E_i$= 45 eV, $E_3$=15 eV